\def\1{\'{\i}}
\def\2{\~n}

\def\ba{\begin{eqnarray}}
\def\ea{\end{eqnarray}}
\def\be{\begin{equation}}
\def\ee{\end{equation}}
\def\beq{\begin{equation}}
\def\eeq{\end{equation}}

\documentclass[preprint,12pt]{elsarticle}

\usepackage{amssymb}
\usepackage{times}
\journal{Annals of Physics}

\begin{document}

\begin{frontmatter}

%% Title, authors and addresses

%% use the tnoteref command within \title for footnotes;
%% use the tnotetext command for the associated footnote;
%% use the fnref command within \author or \address for footnotes;
%% use the fntext command for the associated footnote;
%% use the corref command within \author for corresponding author footnotes;
%% use the cortext command for the associated footnote;
%% use the ead command for the email address,
%% and the form \ead[url] for the home page:
%%
%% \title{Title\tnoteref{label1}}
%% \tnotetext[label1]{}
%% \author{Name\corref{cor1}\fnref{label2}}
%% \ead{email address}
%% \ead[url]{home page}
%% \fntext[label2]{}
%% \cortext[cor1]{}
%% \address{Address\fnref{label3}}
%% \fntext[label3]{}

\title{A short note on \lq\lq Group theoretic approach to rationally extended shape invariant potentials\rq\rq\ [Ann. Phys. 359 (2015) 46--54]}

%% use optional labels to link authors explicitly to addresses:
%% \author[label1,label2]{<author name>}
%% \address[label1]{<address>}
%% \address[label2]{<address>}

\author[add1]{Arturo Ramos}
\ead{aramos@unizar.es}
\author[add2]{Bijan Bagchi}
\ead{bbagchi123@gmail.com}
\author[add3]{Avinash Khare}
\ead{khare@physics.univpune.ac.in }
\author[add4]{Nisha Kumari}
\ead{nishaism0086@gmail.com}
\author[add4]{Bhabani Prasad Mandal}
\ead{bhabani.mandal@gmail.com }
\author[add6]{Rajesh Kumar Yadav}
\ead{rajeshastrophysics@gmail.com}

\address[add1]{Departamento de An\'alisis Econ\'omico,
Universidad de Zaragoza, Gran V\'{\i}a 2, E-50005 Zaragoza, Spain.}
\address[add2]{Department of Physics, School of Natural Science, Shiv Nadar University, Greater Noida-201314, India.}
\address[add3]{Department of Physics, Savitribai Phule Pune University, Pune-411007, India.}
\address[add4]{Department of Physics, Banaras Hindu University, Varanasi-221005, India.}
%\address[add5]{Department of Physics, Banaras Hindu University, Varanasi-221005, India.}
\address[add6]{Department of Physics, S. P. College, S. K. M. University, Dumka-814101, India.}

%\maketitle
 %{$~^a$ Department of Physics, S. P. College, S K. M. University, Dumka-814101, India.\\
 %$~^b$ Department of Physics, Savitribai Phule Pune University, Pune-411007, India.\\
 %$~^c$ Department of Applied Mathematics, University of Calcutta, Kolkata-700 009, India.\\
%$~^d$ Department of Physics, Banaras Hindu University, Varanasi-221005, India.
%}

%\address[labela]{
%Departamento de An\'alisis Econ\'omico,
%Universidad de Zaragoza, \\ Gran V\'{\i}a 2, E-50005 Zaragoza, Spain}
%\ead{aramos@unizar.es}

\begin{abstract}
%% Text of abstract
It is proved the equivalence of the compatibility condition of
[A. Ramos, J. Phys. A 44 (2011) 342001, Phys. Lett. A 376 (2012) 3499] with a condition found in [Yadav et al., Ann. Phys. 359 (2015) 46]. The link of Shape Invariance with the existence of a Potential Algebra is reinforced for the rationally extended Shape Invariant potentials. Some examples on $X_1$ and $X_\ell$ Jacobi and Laguerre cases are given.
\end{abstract}

\begin{keyword}
%% keywords here, in the form: keyword \sep keyword
shape invariance \sep compatibility condition \sep potential algebra

%% MSC codes here, in the form: \MSC code \sep code
%% or \MSC[2008] code \sep code (2000 is the default)
\MSC 81Q05 \sep 81Q60

\end{keyword}

\end{frontmatter}

%%
%% Start line numbering here if you want
%%
% \linenumbers

%% main text

\section{Introduction}

The concept of Shape Invariance has by now a long tradition in Quantum Mechanics, in the search of exactly solvable potentials. It started in the classical work of Infeld and Hull \cite{InfHul51} or even in the seminal works of Schr\"odinger himself \cite{Sch40,Sch41,Sch41b}.
Later, the concept was reformulated by Gendenshte\"{\i}n and Krive \cite{Gen83,GenKri85} as it is known today. See the relatively recent monographes \cite{CooKhaSuk01,GanMalRas11} for an overview. The list of Shape Invariant potentials remained unchanged until
G\'{o}mez-Ullate, Kamran and Milson \cite{GomKamMil09} realized that the classical orthogonal polynomials can be generalized to a situation in which the lowest degree of the polynomials of the family need not to be zero. This fostered the key developments of Quesne and collaborators \cite{Que08,Que09,BagQueRoy09,Que11,Que12,Que12b}, who have shown that it is possible to rationally extend some types of the standard Shape Invariant potentials in order to give isospectral ones.
This line of research has been followed by many authors, for example important contributions by Grandati and collaborators
\cite{Gra11,Gra12,Gra12b,Gra14,GraBer10,GraBer13,GomGraMil14} and by Odake and Sasaki \cite{OdaSas09,OdaSas10,OdaSas10b,OdaSas10c,OdaSas11,OdaSas11b}.
On his side, Ramos \cite{Ram11,Ram12} has found a compatibility condition that the new extended Shape Invariant potentials have to satisfy. Apart {}from that, in \cite{BagQue00} it has been considered the complex Lie algebra $sl(2,\mathbb{C})$ when dealing with non-Hermitian Hamiltonians with real eigenvalues.
Later on, in \cite{YadKumKhaMan15} a group theoretical approach to some extended Shape Invariant potentials has been developed, in which another condition has to be satisfied by the seed superpotential and the functions defining the extension. This technique was further employed by Yadav et al. in \cite{YadKhaBagKumMan16}. However, they did not discuss the relation of their condition with the previously mentioned compatibility condition of \cite{Ram11,Ram12}.

The aim of this short note is to show that the compatibility condition of \cite{Ram11,Ram12} and the condition of \cite{YadKumKhaMan15} are indeed equivalent. This has the important consequence that the former compatibility condition is given in this way a group theoretical sense, and that the overall picture becomes unified in a common setting.
In Section~\ref{equiv} we prove the mentioned equivalence.
In Section~\ref{examples} we provide some examples.
In the final section we provide an Outlook and some Conclusions.

\section{Equivalence of the conditions}\label{equiv}

\subsection{The compatibility condition}\label{ccap}

We briefly recall here the compatibility condition approach of \cite{Ram11,Ram12}.
Given a superpotential of the type
\begin{equation}
W(x,m)=W_0(x,m)+W_{1+}(x,m)-W_{1-}(x,m)
\label{supp}
\end{equation}
where $x$ is the coordinate of the problem under study, $m$ is a parameter that is transformed by translation (by $f(m)=m-1$, without loss of generality), and $W_0(x,m)=k_0(x)+m k_1(x)$ is a superpotential of the affine in $m$ type treated by Infeld and Hull \cite{InfHul51}. $W_{1+}(x,m)$, $W_{1-}(x,m)$ are logarithmic derivatives which moreover satisfy
\begin{equation}
W_{1-}(x,m)=W_{1+}(x,m-1)
\label{peqcc}
\end{equation}
In \cite[Theorem 1]{Ram12} it has been proved that the superpotential (\ref{supp}) defines a Shape Invariant pair of partner potentials through the usual Riccati
equations if and only if it is satisfied
\begin{eqnarray}
& &W^2_{1+}(x,m)+W^\prime_{1+}(x,m)+W^2_{1-}(x,m)
+W^\prime_{1-}(x,m)\nonumber\\
& &-2 W_0(x,m)W_{1-}(x,m)\nonumber\\
& &+2 W_0(x,m)W_{1+}(x,m)-2W_{1-}(x,m)W_{1+}(x,m)=\epsilon(x)\label{cc}
\end{eqnarray}
where $\epsilon(x)$ is a function of $x$ only.
Since (\ref{cc}) holds for all allowed $m$'s, in particular it holds as well for $m-1$.

\subsection{The group theory approach}\label{gtapp}

On its side, Yadav et al. \cite{YadKumKhaMan15} have developed a group theoretical approach to some rationally extended Shape Invariant potentials. They were inspired by the well-known paper of Wu and Alhassid \cite{WuAlh90} and in this way not all the possible cases of affine in $m$ Shape Invariant potentials are considered.
Therefore, in this paper we consider a slightly more general approach inspired by Miller \cite{Mil68} combined with the previous two papers.

That is, we now consider the $G(a,b)$ Potential Algebra
by means of the operators
\begin{eqnarray}
J_{\pm}&=&{\rm e}^{\pm i\phi}\left[\pm\frac{\partial}{\partial x}
-\left(\left(-i\frac{\partial}{\partial\phi}\pm\frac 1 2\right)F(x)-G(x)\right)\right.\nonumber\\
& &\left.\quad\quad -U\left(x,-i\frac{\partial}{\partial\phi}\pm\frac 1 2\right)
\right]\label{defJpm}\\
J_3&=&-i\frac{\partial}{\partial\phi} \label{defJ3}\\
E&=&1\label{defE}
\end{eqnarray}
where $F(x), G(x)$ are functions of $x$ only, $a$ and $b$ are real numbers, and $U\left(x,-i\frac{\partial}{\partial\phi}\pm\frac 1 2\right)$
is a functional operator. The commutation relations
\begin{eqnarray}
& &[J_+,E]=0\,,\quad [J_-,E]=0\,,\quad [J_3,E]=0\,,\label{comm1}\\
& &[J_3,J_+]=J_+\,,\quad [J_3,J_-]=-J_-\,,\quad\label{comm2}
\end{eqnarray}
are satisfied automatically.
The commutation relation
\begin{equation}
[J_+,J_-]=-2 a J_3-2 b E
\label{comm3}
\end{equation}
is satisfied (when evaluated in a basis of eigenfunctions of
the Casimir \cite{Mil68} $C=a J_{3}^2-a J_{3}+2 b J_{3}E-J_{+}J_{-}=a J_{3}^2+a J_{3}+2 b J_{3}E+2bE-J_{-}J_{+}$ and $J_3$, $\psi_{cm}(x){\rm e}^{i m\phi}$) if and only if three conditions do. The first two are analogous to the conditions of \cite{Mil68}, and the third one is an additional condition. They are:
\begin{equation}
F^\prime(x)+F^2(x)=a\,,\quad G^\prime(x)+F(x)G(x)=b
\label{ih}
\end{equation}
and
\begin{eqnarray}
&&U^2\left(x,m-\frac 1 2\right)
-U^\prime\left(x,m-\frac 1 2\right)\nonumber\\
&&+2U\left(x,m-\frac 1 2\right)\left(F(x)\left(m-\frac 1 2\right)-G(x)\right)\nonumber\\
&&-U^2\left(x,m+\frac 1 2\right)
-U^\prime\left(x,m+\frac 1 2\right)\nonumber\\
&&-2U\left(x,m+\frac 1 2\right)\left(F(x)\left(m+\frac 1 2\right)-G(x)\right)=0\label{ccin}
\end{eqnarray}

\subsection{Relation between the two approaches}

The aim of this note is to establish the relation between (\ref{ccin}) and (\ref{peqcc}), (\ref{cc}). So let us
start {}from the left hand side of relation (\ref{ccin}). Performing a change of parameter $m$, without loss of generality, $m\rightarrow m-\frac 1 2$, we obtain
\begin{eqnarray}
& &U^2(x,m-1)-2G(x)(U(x,m-1)-U(x,m))-U^2(x,m)\nonumber\\
& &+2F(x)((m-1)U(x,m-1)-m U(x,m))\nonumber\\
& &-U^\prime(x,m-1)-U^\prime(x,m)\label{step1}
\end{eqnarray}
Then, linking the notations in the Subsection~\ref{ccap} with those of Subsection~\ref{gtapp} in the following way (see eqs. (34) and (35) in \cite{YadKumKhaMan15})
\begin{eqnarray}
&& F(x)=k_1(x)\label{identF}\\
&& G(x)=-k_0(x)\label{identG}\\
&& U(x,m)=W_{1+}(x,m)-W_{1-}(x,m)\label{identUm}\\
&& U(x,m-1)=W_{1+}(x,m-1)-W_{1-}(x,m-1)\label{identUmm1}
\end{eqnarray}
we obtain
\begin{eqnarray}
&& -2(k_0(x)+(m-1)k_1(x))(W_{1-}(x,m-1)-W_{1+}(x,m-1))\nonumber\\
&&+(W_{1-}(x,m-1)-W_{1+}(x,m-1))^2\nonumber\\
&& +2(k_0(x)+m k_1(x))(W_{1-}(x,m)-W_{1+}(x,m))
-(W_{1-}(x,m)-W_{1+}(x,m))^2\nonumber\\
&&+W^\prime_{1-}(x,m-1)+W^\prime_{1-}(x,m)
-W^\prime_{1+}(x,m-1)-W^\prime_{1+}(x,m)\label{step2}
\end{eqnarray}
Then, using (\ref{cc}) for $m$ and $m-1$, the expression simply reduces to
\begin{equation}
-2 W^\prime_{1+}(x,m-1)+2 W^\prime_{1-}(x,m)
\label{step3}
\end{equation}
which, by virtue of (\ref{peqcc}), vanishes identically.
The steps can be reversed in a natural way so it is proved the equivalence of (\ref{peqcc}), (\ref{cc}) with (\ref{ccin}).

\section{Examples}\label{examples}

In this Section we illustrate the applicability of the previous
relation in several instances, including extensions of Shape Invariant potentials with $X_1$ and $X_\ell$ Jacobi and Laguerre polynomials.

\subsection{Case of $k_0(x)+m k_1(x)=-\frac{\beta}{c} \coth(c x) + \frac{d}{\sinh(c x)}+m c \coth(c x)$, $X_1$ extension}

For this case $k_0(x)=-G(x)=-\frac{\beta}{c} \coth(c x) + \frac{d}{\sinh(c x)}$ and $k_1(x)=F(x)=c \coth(c x)$ where $x\in(0,\infty)$, $c>0$, $\beta,d$ are constants. This example is a slight generalization of one appeared first in \cite{Que09,BagQueRoy09} and is a slight correction of one appeared in \cite{Ram11}. We can take
\begin{eqnarray}
W_{1+}(x,m)&=&\frac{2 c^2 d \sinh(c x)}
{-2\beta+c^2(2m+1)+2 c d \cosh(cx)}\label{ex1Wp}\\
W_{1-}(x,m)&=&\frac{2 c^2 d \sinh(c x)}
{-2\beta+c^2(2m-1)+2 c d \cosh(cx)}\label{ex1Wm}
\end{eqnarray}
and then with the identifications above it is readily checked that (\ref{peqcc}), (\ref{cc}) are satisfied. Likewise, (\ref{ih})
is satisfied with $a=c^2$ and $b=\beta$, and (\ref{ccin}) is also satisfied. This example leads to a pair of Shape Invariant partner potentials which are non-singular if
$\beta$ is real, $d<0$ and $m<\frac{2\beta-c^2-2cd}{2 c^2}$ or if
$d>0$ when $m>\frac{2\beta+c^2-2cd}{2 c^2}$.

\subsection{Case of $k_0(x)+m k_1(x)=\frac{\omega x}{2}+\frac{d}{x}+\frac{m}{x}$, radial oscillator
with $X_1$ extension}

For this case $k_0(x)=-G(x)=\frac{\omega x}{2}+\frac{d}{x}$ and $k_1(x)=F(x)=\frac{1}{x}$ where $x\in(0,\infty)$, and $\omega>0$, $d>0$ are two constants. This example is a slight generalization of one appeared first in \cite{Que08} and is a modification of one appeared in \cite{Ram11}. We can take
\begin{eqnarray}
W_{1+}(x,m)&=&-\frac{2\omega x}
{1+2d+2m-\omega x^2}\label{ex2Wp}\\
W_{1-}(x,m)&=&-\frac{2\omega x}
{-1+2d+2m-\omega x^2}\label{ex2Wm}
\end{eqnarray}
and then with the identifications above it is readily checked that (\ref{peqcc}), (\ref{cc}) are satisfied. Likewise, (\ref{ih})
is satisfied with $a=0$ and $b=-\omega$, and (\ref{ccin}) is also satisfied. This example leads to a pair of Shape Invariant partner potentials which are non-singular if $m<-\frac 1 2(1+2d)$.

\subsection{Case of $k_0(x)+m k_1(x)=-\frac{\beta}{c} \tan(c x) + \frac{d}{\cos(c x)}-m c \tan(c x)$,$X_1$ extension}

For this case $k_0(x)=-G(x)=-\frac{\beta}{c}\tan(cx)+\frac{d}{\cos(cx)}$ and $k_1(x)=F(x)=-c\tan(cx)$ where $x\in\left(-\frac{\pi}{2c},\frac{\pi}{2c}\right)$, $c>0$, $\beta,d$ are constants. This example is a slight generalization of one appeared first in \cite{Que08} and is a modification of one appeared in \cite{Ram11}. We can take
\begin{eqnarray}
W_{1+}(x,m)&=&-\frac{2 c^2 d \cos(c x)}
{2\beta+c^2(1+2m)-2 c d \sin(cx)}\label{ex3Wp}\\
W_{1-}(x,m)&=&\frac{2 c^2 d \cos(c x)}
{-2\beta+c^2(1-2m)+2 c d \sin(cx)}\label{ex3Wm}
\end{eqnarray}
and then with the identifications above it is readily checked that (\ref{peqcc}), (\ref{cc}) are satisfied. Likewise, (\ref{ih})
is satisfied with $a=-c^2$ and $b=\beta$, and (\ref{ccin}) is also satisfied. This example leads to a pair of Shape Invariant partner potentials which are non-singular if
$\beta$ is real, $d>0$ and $m<\frac{-2\beta-c^2-2cd}{2 c^2}$ or if
$d<0$ when $m>\frac{2\beta+c^2-2cd}{2 c^2}$, or also if $\beta$ is real, $d>0$ and $m>\frac{-2\beta+c^2+2cd}{2 c^2}$ or if
$d<0$ when $m<\frac{-2\beta-c^2+2cd}{2 c^2}$.

 \subsection{Case of $k_0(x)+m k_1(x)=-\frac{B}{\sinh(x)}+m\coth(x)$, Generalized P\"osch--Teller potential with $X_\ell$ extension}

For this case $k_0(x)=-G(x)=-\frac{B}{\sinh(x)}$ and $k_1(x)=F(x)=\coth(x)$ where $x\in(0,\infty)$, $B$ is a real constant. This example appeared in \cite{YadKumKhaMan15}, inspired in \cite{Que09,BagQueRoy09}. We can take
\begin{eqnarray}
W_{1+}(x,\ell,m)&=&\frac 1 2(\ell-2B-1)\sinh(x)
\frac{P_{\ell-1}^{(-B+m+1/2,-B-m-1/2)}(\cosh(x))}
{P_{\ell}^{(-B+m-1/2,-B-m-3/2)}(\cosh(x))}\nonumber\\
& &\label{ex4Wp}\\
W_{1-}(x,\ell,m)&=&\frac 1 2(\ell-2B-1)\sinh(x)
\frac{P_{\ell-1}^{(-B+m-1/2,-B-m+1/2)}(\cosh(x))}
{P_{\ell}^{(-B+m-3/2,-B-m-1/2)}(\cosh(x))}\nonumber\\
& &\label{ex4Wm}
\end{eqnarray}
where $P_\ell^{(\alpha,\beta)}(x)$ denotes the ordinary $\ell$-th Jacobi polynomial.
Then, with the identifications above it is readily checked that (\ref{peqcc}), (\ref{cc}) are satisfied. Likewise, (\ref{ih})
is satisfied with $a=1$ and $b=0$, and (\ref{ccin}) is also satisfied. This example leads to a pair of Shape Invariant partner potentials which are non-singular if $B<-\frac{1}{2}$ and $\frac{1}{2}(1+2B)<m<-\frac{1}{2}(1+2B)$ (with these conditions it is ensured that the roots of the Jacobi polynomials in the denominators above are on the interval $(-1,1)$ and then $\cosh(x)$ takes values in $[1,\infty)$).

\subsection{Case of $k_0(x)+m k_1(x)=\frac{i B}{\cosh(x)}+m\tanh(x)$, PT symmetric complex Scarf-II with $X_\ell$ extension}

For this case $k_0(x)=-G(x)=\frac{iB}{\cosh(x)}$ and $k_1(x)=F(x)=\tanh(x)$ where $x\in(-\infty,\infty)$, $B$ is a real constant and $i$ is the imaginary unit. This example appeared in \cite{YadKumKhaMan15}, inspired in \cite{Que09,BagQueRoy09,BagQue00,BagQue02}. We can take
\begin{eqnarray}
W_{1+}(x,\ell,m)&=&\frac 1 2 i(\ell-2B-1)\cosh(x)
\frac{P_{\ell-1}^{(-B+m+1/2,-B-m-1/2)}(i\sinh(x))}
{P_{\ell}^{(-B+m-1/2,-B-m-3/2)}(i\sinh(x))}\nonumber\\
& &
\label{ex5Wp}\\
W_{1-}(x,\ell,m)&=&\frac 1 2 i(\ell-2B-1)\cosh(x)
\frac{P_{\ell-1}^{(-B+m-1/2,-B-m+1/2)}(i \sinh(x))}
{P_{\ell}^{(-B+m-3/2,-B-m-1/2)}(i \sinh(x))}
\nonumber\\
& &\label{ex5Wm}
\end{eqnarray}
Then, with the identifications above it is readily checked that (\ref{peqcc}), (\ref{cc}) are satisfied. Likewise, (\ref{ih})
is satisfied with $a=1$ and $b=0$, and (\ref{ccin}) is also satisfied. This example leads to a pair of Shape Invariant partner potentials which are non-singular (except maybe at $x=0$) because the argument of the Jacobi polynomials above is purely imaginary.

\subsection{Case of $k_0(x)+m k_1(x)=\frac{\omega x}{2}+\frac{m}{x}$, radial oscillator with $X_\ell$ extension}

For this case $k_0(x)=-G(x)=\frac{\omega x}{2}$ and $k_1(x)=F(x)=\frac{1}{x}$ where $x\in(0,\infty)$ and $\omega>0$. This example is a slight modification of one appeared first in \cite{Que08,OdaSas09}.
We can take
\begin{eqnarray}
W_{1+}(x,\ell,m)&=&\omega x\frac{L_{\ell-1}^{(-m-1/2)}\left(-\frac{\omega x^2}{2}\right)}
{L_{\ell}^{(-m-3/2)}\left(-\frac{\omega x^2}{2}\right)}
\label{ex6Wp}\\
W_{1-}(x,\ell,m)&=&\omega x\frac{L_{\ell-1}^{(-m+1/2)}\left(-\frac{\omega x^2}{2}\right)}
{L_{\ell}^{(-m-1/2)}\left(-\frac{\omega x^2}{2}\right)}
\label{ex6Wm}
\end{eqnarray}
where now $L_\ell^{(\alpha)}(x)$ denotes the $\ell$-th (associated) Laguerre polynomial. With the identifications above it is readily checked that (\ref{peqcc}), (\ref{cc}) are satisfied. Likewise, (\ref{ih})
is satisfied with $a=0$ and $b=-\omega$, and (\ref{ccin}) is also satisfied. This example leads to a pair of Shape Invariant partner potentials which are non-singular since the roots of the Laguerre polynomials of the denominators above lie in $(0,\infty)$, and we have taken explicitly negative arguments in them by means of
$-\frac{\omega x^2}{2}$, see also \cite{OdaSas09}.

\section{Conclusions and outlook}

We have demonstrated, in a general way and by means of examples,
the validity of the equivalence of the compatibility conditions
(\ref{peqcc}), (\ref{cc}) with the group theoretical condition (\ref{ccin}). Thus both approaches are linked in a clear way. The first two relations establish the Shape Invariance condition of the by now well-known rationally extended potentials, and the last one is one of the conditions for the closing of the extended potential algebra $G(a,b)$, inspired by Miller \cite{Mil68}.
Thus extended Shape Invariance is linked with the closing of a Potential Lie Algebra, initially being an approach known at least since the works \cite{Mil68,WuAlh90} for some of the classical cases of Infeld and Hull \cite{InfHul51}. Thus the classical results are shown to be valid in a new situation.
As a possible extension of the methods employed here, we could
try to model rationally extended Shape Invariant potentials with two \cite{OdaSas09,OdaSas10,OdaSas10b} or more parameters subject to translation with a Potential Algebra, using perhaps the insight of \cite{CarRam00}. This is work to be done in another paper(s).

\section*{Acknowledgements}
The work of A.R. is supported by Aragon Government,
ADETRE Consolidated Group. A.K. is grateful to Indian National
Science Academy (INSA) for  awarding him INSA Senior
Scientist position at Savitribai Phule Pune University.

%\section*{References}

%% The Appendices part is started with the command \appendix;
%% appendix sections are then done as normal sections
%% \appendix

%% \section{}
%% \label{}

%% References
%%
%% Following citation commands can be used in the body text:
%% Usage of \cite is as follows:
%%   \cite{key}         ==>>  [#]
%%   \cite[chap. 2]{key} ==>> [#, chap. 2]
%%

%% References with bibTeX database:

%\bibliographystyle{elsarticle-num}
%\bibliography{<your-bib-database>}

%% Authors are advised to submit their bibtex database files. They are
%% requested to list a bibtex style file in the manuscript if they do
%% not want to use elsarticle-num.bst.

%% References without bibTeX database:

% \begin{thebibliography}{00}

%% \bibitem must have the following form:
%%   \bibitem{key}...
%%

% \bibitem{}

% \end{thebibliography}

\end{document}